# Open bibliographic data and the Italian National Scientific Qualification: measuring coverage of academic fields

Federica Bologna[1], Angelo Di Iorio[2], Silvio Peroni[3,4] and Francesco Poggi[5,6]

[1] Bowers College of Computing and Information Science, Cornell University, Ithaca, USA
[2] Department of Computer Science and Engineering, University of Bologna, Bologna, Italy
[3] Research Centre for Open Scholarly Metadata, Department of Classical Philology and Italian Studies, University of Bologna, Bologna, Italy
[4] Digital Humanities Advanced Research Centre (/DH.arc), Department of Classical Philology and Italian Studies, University of Bologna, Bologna, Italy
[5] Department of Communication and Economics, University of Modena and Reggio Emilia, Reggio Emilia, Italy
[6] Institute of Cognitive Sciences and Technologies, Italian National Research Council (CNR), Rome, Italy

## Author Note

Federica Bologna  https://orcid.org/0000-0002-3845-8266

Angelo Di Iorio  https://orcid.org/0000-0002-6893-7452

Silvio Peroni  https://orcid.org/0000-0003-0530-4305

Francesco Poggi  https://orcid.org/0000-0001-6577-5606

Data and software developed for this work are available at https://doi.org/10.5281/ZENODO.5025114 and in the Github repository at https://github.com/sosgang/coverage_asn. We have no conflicts of interest to disclose. We gratefully acknowledge support from the University fund for Research 2020 (FAR) of the University of Modena and Reggio Emilia, and from the European Union's Horizon 2020 research and innovation program under grant agreement No. 101017452.

Correspondence concerning this article should be addressed to Silvio Peroni: silvio.peroni@unibo.it.

**Abstract**

The importance of open bibliographic repositories is widely accepted by the scientific community. For evaluation processes, however, there is still some skepticism: even if large repositories of open access articles and free publication indexes exist and are continuously growing, assessment procedures still rely on proprietary databases, mainly due to the richness of the data available in these proprietary databases and the services provided by the companies they are offered by. This paper investigates the status of open bibliographic data of three of the most used open resources, namely Microsoft Academic Graph, Crossref and OpenAIRE, evaluating their potentialities as substitutes of proprietary databases for academic evaluation processes. We focused on the Italian National Scientific Qualification (NSQ), the Italian process for University Professor qualification, which uses data from commercial indexes, and investigated similarities and differences between research areas, disciplines and application roles. The main conclusion is that open datasets are ready to be used for some disciplines, among which mathematics, natural sciences, economics and statistics, even if there is still room for improvement; but there is still a large gap to fill in others - like history, philosophy, pedagogy and psychology - and a stronger effort is required from researchers and institutions.

## 1. Introduction

The relevance of open bibliographic datasets is continuously increasing, under the push of internationally-coordinated efforts such as the Initiative for Open Citations (I4OC, https://i4oc.org) and the Initiative for Open Abstracts (I4OA, https://i4oa.org). These datasets not only changed the way scholars search for literature, but also enabled the advancement of the scientometrics and bibliometrics fields. The larger availability of open data made it possible for researchers to carry out ground-breaking studies on research practices, academic literature and academic institutions (Bedogni et al., 2021; Chudlarský & Dvořák, 2020; Di Iorio et al., 2019; Huang et al., 2020; Martín-Martín et al., 2021; S. Peroni et al., 2020; Zhu et al., 2020).

One of the issues still open in this context is whether or not the open bibliographic datasets are ready to substitute commercial ones in the research evaluation processes. In fact, bibliometrics are being widely used, both by private and governmental agencies, to evaluate the scientific performance of institutions, journals, groups and scholars. For example, several countries employ evaluation procedures that combine bibliometrics and peer-review, like the *Excellence in Research for Australia* (ERA), The British *Research Excellence Framework* (REF), the *Valutazione della Qualità della Ricerca* (VQR) and the *National Scientific Qualification* (NSQ) in Italy. But these processes still rely on commercial datasets, such as Scopus and WoS.

Therefore investigating the availability of open data in the context of national scholarly assessments is of utmost relevance. The first step in that direction is to study the coverage of the publications in these datasets. Previous works have mainly used (the publications found in) proprietary datasets as benchmarks against which to compare (the publications found in) open datasets (Martín-Martín et al. 2021; Huang et al. 2020; Singh et al. 2021; Visser et al. 2021; Harzing 2019).

This paper adds a piece to the puzzle. It sheds light on the differences between the open datasets when used for evaluating research productivity in different contexts and disciplines.

Specifically, we ground this study in the Italian National Scientific Qualification (NSQ). The NSQ is a nation-wide research assessment exercise which establishes whether a scholar can apply to professorial academic positions as Associate Professor and Full Professor. Applications are organized according to a governmentally-defined taxonomy of 190 Recruitment Fields (RF) divided into 14 Scientific Areas (SA). The disciplines are divided into two categories, citation-based (CDs) and non-citation-based (NDs), depending on the use of citations for the evaluation. In the NSQ nomenclature, CDs and NDs are actually tagged as "bibliometric" and "non-bibliometric" disciplines, respectively, but we prefer not to use this terminology that might be misleading here. Note also that SAs can include either CDs or NDs disciplines or a mix of them.

Our work moves off such a distinction and digs into open datasets: do these datasets show the same structure and behavior for CDs and NDs disciplines? Which are the most relevant differences? Where do these differences derive from? And more: are there significant differences between the coverage for candidates as Associate Professor or Full Professor? And between the coverage for candidates in different recruitment fields?

To answer these questions we analyzed Microsoft Academic Graph, Crossref, and OpenAIRE and compared the coverage of publications for a wide range of disciplines and candidates. This work in fact is an extension of a particular aspect (i.e. the coverage of publications in open datasets) of our previous study that was limited to some recruitment fields only (Bologna et al, 2021c). Here, we include all disciplines of the NSQ and analyze the publications of all candidates in the 2016, 2017, and 2018 terms of the NSQ. Overall we consider 2,353,872 publications for 58,335 candidates in 190 Recruitment Fields. We also investigate whether coverage improves when combining the three datasets, collecting evidence of the *effectiveness* of these datasets for CDs disciplines, much more than NDs, and some peculiarities of each dataset.

Note that the term *effectiveness* is used here to indicate the capability of providing data for the evaluation process, not the fairness and efficacy of the process itself. Our goal is not to assess the NSQ or similar processes, rather to understand if these processes could be built and improved by using open data only. Having good coverage is a necessary but not sufficient condition for such a transition. The goal here is to assess the data collection process, understanding how much information is available today, from which sources and for which disciplines.

## 2. Related works

*2.1. Previous works on coverage*

Numerous studies analyzing coverage of bibliographic datasets have been published since these became widely used in the scientific community. These studies differ in the methods and measurements they use to analyze and compare coverage, in the sets of datasets they compare and on the type of document they focus their comparison on.

Some studies focus on a small sample of documents belonging to a specific discipline or single author (Harzing, 2019). Others involve millions of documents from a wide range of disciplines (Martín-Martín et al., 2021). Some works employ a straightforward method, obtaining the complete list of documents contained in each dataset, matching the documents across sources and measuring the overlap (Visser et al., 2021). Other works use alternative methods, since not all bibliographic datasets offer free access to their data, by comparing documents' citation lists (Martín-Martín et al., 2021; Martín-Martín et al., 2018).

Of these previous analyses on coverage in bibliographic datasets, a great number focus on WoS and Scopus, and use them as benchmarks to evaluate other datasets' coverage. Some studies draw comparisons in coverage between WoS and Scopus (Mongeon and Paul-Hus, 2016). Some compare their coverage to that of Dimensions (Singh et al., 2021; Thelwall, 2018; Orduña-Malea & Delgado-López-Cózar, 2018), of Google Scholar (Delgado López-Cózar et al., 2019; Martín-Martín et al., 2018), of Microsoft Academic (Huang et al., 2020; Hug & Brändle, 2017), of both Google Scholar and Microsoft Academic (Harzing 2016; Harzing & Alakangas, 2017a, 2017b). Others compare multiple datasets against each other (Martín-Martín et al., 2021;

Visser et al., 2021; Harzing 2019).

*2.2. Bibliographic Datasets*

For more than a decade, Web of Science (WoS, introduced online in 1997) (Birkle et. al, 2020) and Scopus (launched in 2004) (Baas et al., 2020) have been the only available options to conduct large-scale bibliometric analyses. These two commercial subscription-based bibliographic data sources provide metadata on scientific documents and on citation links between these documents. Google Scholars (Van Noorden, 2014), a free bibliographic search engine, was launched a week after Scopus. However, it did not provide bulk-access to its data.

The introduction of new open bibliographic data sources changed this trend (Martín-Martín et al., 2021). In 2013, Crossref, a nonprofit membership association between publishers, made all its metadata available to the public via a REST API, and in 2017, thanks to the Initiative for Open Citations (I4OC; https://i4oc.org), millions of citation links between documents have also been made openly available. In 2014, OpenAIRE (Manghi et al., 2012), an EU-funded infrastructure to share bibliographic data across institutions, released free API access to its data. In 2016, Microsoft Academic (Wang et al. 2020) provided a scholarly search engine and bulk access to its data via an API, both without charge. This study focuses on these mentioned open-access datasets.

A number of other bibliographic data sources have not been considered in this study, for various reasons:
- Dimensions (Herzog et al. 2020) requires payment of a fee for bulk access;
- CiteSeerX (Wu et al. 2019) indexes documents in the public web, but not those behind paywalls;
- ResearchGate (https://www.researchgate.net/) does not provide a tool to extract data in bulk;
- Lens.org provides free bulk-access to non-commercial projects only for a limited time and

   at a limited access rate;
- Regional- and subject-specific datasets do not offer multidisciplinary coverage by design; hence, they are not comparable to the other sources considered in this study.

*2.3. The Italian National Scientific Qualification (NSQ)*

In 2011, Italian Law of December 30th 2010 n.240 (L. 240/2010, 2011) implemented the NSQ, a nation-wide research assessment exercise that attests the scientific maturity of scholars. The law made it mandatory to pass the NSQ in order to apply to academic positions. The NSQ consists of two distinct qualification processes, one for the academic position of Full Professor (FP), and one for that of Associate Professor (AP). Passing the NSQ does not grant a tenure position. It is each university's responsibility to create new positions and hire scholars according to financial and administrative requirements.

Moreover, the Ministerial Decree of June 14th 2012 (D.L. 2012, 2012) defined a taxonomy of 184 (extended to 190 a few years later) Recruitment Fields (RF) divided into groups and sorted into 14 different Scientific Areas (SA). SAs correspond to vast academic disciplines, whereas RFs correspond to specific scientific fields of study. Each scholar is assigned to a specific RF which belongs to a single SA. In the taxonomy, RFs are identified by an alphanumeric code in the form AA/GF. AA is a number indicating the SA, ranging from 1 to 14. G is a single letter identifying the group of RFs. F is a digit indicating the RF. For instance, Neurology's code is 06/D5, where 06 indicates the SA Medicine and D indicates the group Specialized Clinical Medicine (D.L. 2012, 2012). When applying for the NSQ, scholars can choose to be evaluated for more RFs at a time. Since each RF has its own assessment rules, the candidate may pass the qualification in some fields but not in others.

The NSQ divides academic disciplines into two categories, i.e. citation-based disciplines (CDs) and non-citation-based disciplines (NDs). This division affects only the metrics used for assessing the candidates of that discipline in the first part of the process. Candidates applying to

CDs are evaluated using:

- CD_M1: the number of their journal papers;
- CD_M2: the total number of citations received;
- CD_M3: their h-index.

While candidates applying to NDs are evaluated using:

- ND_M1: number of their journal papers and book chapters;
- ND_M2: number of their papers published on Class A journals[1];
- ND_M3: number of their published books.

In order to apply to the NSQ, candidates have to submit a curriculum vitae (CV) with detailed information about their research accomplishments. Then, NSQ assessment is organized in two steps. In the first step of the evaluation, candidates' metrics are expected to exceed two out of the three thresholds in their RF. Successively, the candidate's maturity is evaluated based on their CV. The aforementioned metrics are computed for each candidate, taking into consideration only publications that are less than 15 years old for candidates to the role of FP and 10 years old for candidates to the role of AP. This process utilizes data retrieved from Scopus and Web of Science and is conducted by the *Italian National Agency for the Evaluation of Universities and Research Institutes* (ANVUR). ANVUR also sets thresholds for each metric by RF. Normalization based on the scholars' scientific age (the number of years since the first publication) is used to compute most of the metrics.

As shown in Table 1, citation-based disciplines are predominantly either STEM-based (Science, technology, engineering, and mathematics) or Medicine-based RFs, such as all the RFs in the first nine SAs (01-09), with the exception of the RFs 08/C1, 08/D1, 08/E1, 08/E2, 08/F1, which are considered NDs, and the four RFs in Psychology (11/E), which are considered CDs. Whereas non-citation-based disciplines are predominantly HASS-based (Humanities, Arts and Social Sciences) RFs, such as the last five SAs (10-14) with the exceptions just described. The reason for this division is that, according to ANVUR, reliable and sufficiently complete citation databases exist for CDs, but they do not for NDs.

---

[1] The top-rated journals according to official classification provided by ANVUR available at https://www.anvur.it/attivita/classificazione-delle-riviste/classificazione-delle-riviste-ai-fini-dellabilitazione-scientifica-nazionale/elenchi-di-riviste-scientifiche-e-di-classe-a/.

|      | citation-based disciplines (CDs)                          | non-citation-based disciplines (NDs)                                                  |
| ---- | --------------------------------------------------------- | ------------------------------------------------------------------------------------- |
| SA01 | all RFs                                                   | no RFs                                                                                |
| SA02 | all RFs                                                   | no RFs                                                                                |
| SA03 | all RFs                                                   | no RFs                                                                                |
| SA04 | all RFs                                                   | no RFs                                                                                |
| SA05 | all RFs                                                   | no RFs                                                                                |
| SA06 | all RFs                                                   | no RFs                                                                                |
| SA07 | all RFs                                                   | no RFs                                                                                |
| SA08 | RFs 08/A1, 08/A2, 08/A3, 08/A4, 08/B1, 08/B2, 08/B3       | RFs 08/C1, 08/D1, 08/E1, 08/E2, 08/F1                                                 |
| SA09 | all RFs                                                   | no RFs                                                                                |
| SA10 | no RFs                                                    | all RFs                                                                               |
| SA11 | RFs 11/E1, 11/E2, 11/E3, 11/E4                            | RFs 11/A1, 11/A2, 11/A3, 11/A4, 11/A5, 11/B1, 11/C1, 11/C2, 11/C3, 11/C4, 11/C5, 11/D1 |
| SA12 | no RFs                                                    | all RFs                                                                               |
| SA13 | no RFs                                                    | all RFs                                                                               |
| SA14 | no RFs                                                    | all RFs                                                                               |

*Table 1. Identification of the RFs included in the various SAs that are defined either as citation-based disciplines (CDs) or non-citation-based disciplines (NDs) according to ANVUR.*

For the purposes of this study, we take into consideration the second session of the NSQ

which took place from 2016 to 2018: with 1 term in 2016, 2 terms in 2017 and 2 terms in 2018.

In order to keep the process as transparent as possible, the full CV of each candidate is publicly posted (in PDF format) on the NSQ website and is accompanied by the full scripts of the judgements of the evaluation committee (a.k.a. a commission) of each RF – composed of five full professors responsible for assessing applicants for AP and FP. We use the metadata contained in these CVs to investigate and compare coverage of the candidates' publications by Microsoft Academic Graph, Crossref and OpenAIRE across disciplines.

## 3. Methods and materials

This section introduces all the methods and materials used for our study. Data and software developed for this work are available in (Bologna et al, 2021b) and in the project Github repository https://github.com/sosgang/coverage_asn.

| | |
|---|---|
| 58,364 | unique applications considered |
| 58,335 | unique applications with relevant publishing data, of which 41,668 applications to CDs and 16,668 applications to NDs |
| 9 | missing CVs |
| 19 | CVs without relevant publishing data |
| 2,353,872 | publications with metadata, of which 1,951,515 publications in CDs 402,357 publications in NDs |
| 17 | publications without any metadata |
| 18437 | publications with parsing issues |
| 2384 | publications without enough metadata |

*Table 2. Number of applications and publications considered in the study.*

*3.1. Data*

For the purposes of this study, we considered all candidates that participated in the 2016, 2017, and 2018 sessions of the NSQ. From each candidate's CV, we extracted all the available publications metadata as described in the following section. The specifics of the dataset obtained from these CVs are available in Table 2.

*3.2. Sources*

We considered three open access sources. The first, Microsoft Academic Graph (https://www.microsoft.com/en-us/research/project/microsoft-academic-graph/) (Wang et al., 2020), referred to as MAG here, results from the efforts of the Microsoft Academic Search (MAS) project. This dataset is updated biweekly and is distributed under an open data license for research and commercial applications. We use a copy of MAG created and made available by Internet Archive in January 2020 (Microsoft Academic, 2020).

The second, OpenAIRE Graph (https://www.openaire.eu/) (Manghi et al., 2012), referred to as OA here, includes information about objects of the scholarly communication life-cycle (publications, research data, research software, projects, organizations, etc.) and semantic links among them. It is created bi-monthly and is accessible for scholarly communication and research analytics. We use the dump that OpenAIRE released on Zenodo in April 2021 (Manghi et al., 2021).

The third, Crossref (https://www.crossref.org/) (Hendricks et al., 2020), referred to as CR here, was born as a nonprofit membership association among publishers to promote collaboration to speed research and innovation. The dataset is fully curated and governed by the members. We use the dump released in January 2021 (Crossref, 2021).

*3.3. Dumps processing and database creation*

In order to efficiently query MAG, OA and CR for each publication of each candidate in each quick succession, we decide to create a database containing all the bibliographic metadata present in each dataset dump.

First, we download and process each dump. We take each publication in the dump, select the metadata of interest for our analysis (author, title, year, DOI, and any MAG-specific identifier) and store it into a JSON file. Thus transforming the three dumps into three large JSON files.

Secondly, we set up the MongoDB database. MAG's, OA's and CR's JSON files are imported as separate collections into a MongoDB database. We then create indexes in each collection to improve query efficiency. We set a compound index, combining a text index on the field "title" of the publications and an ascending index on the field "year", and an ascending index on the field "doi". In MAG's collection we set two other ascending indexes, one on the field "id.mag", containing MAG's publication identifier, and on field "authors.id.mag", containing MAG's author identifiers.

We then proceed to query the database with the candidates' publication metadata to collect coverage information.

*3.4. Querying the database and collecting coverage information*

To obtain coverage information on the candidates' publications, we first extract all bibliographic metadata (e.g. the title, authors and DOIs of the publications) from the CVs the candidates submitted when applying to the NSQ (in the 2016, 2017, and 2018 sessions). The CVs were available in PDF and have been converted into a pure textual format to extract structured information (such as the title, authors and DOIs of the publications) to be stored in JSON.

We obtain a list of publications for each candidate with their relevant metadata (*title*, *year*,

*doi*, *authors*). Then, for each candidate, we use this metadata to query MAG, OA and CR collections in the database to find each publication in each collection. We query the database either by *doi*, if present among the publication's metadata, or by *year* and *title*. In OA, when we find a publication, we add a "coverage marker" to the publication's metadata to signal that said publication is present in the collection. The same is done for CR collection. In MAG, when we find a publication in the database, we collect and store the Paper Id (*PId*), which identifies the publication, and the *Author Id* of the candidate, which identifies the author. The *PId* acts as "coverage marker" for MAG collection. Since in MAG each author can be assigned multiple *Author Ids*, we retrieve one for each publication we find and keep only the unique ids. Then, we query MAG by each *Author Id*, retrieve all publications associated with that id, and compare them to our list of publications. We do so to catch publications that are present in MAG's collection but that we were not able to find by querying the collection using the publication's metadata in the CV.

Lastly, for each candidate, we calculate the coverage of their publications by MAG, OA, CR or the combination of the three. We count the number of publications in the candidate CV for which we have either the *doi*, or *year* and *title* information. We then count the number of publications that have a *PId*, the number of publications that have OA's coverage marker, and the number of publications that have CR's coverage marker, and the number of publications that have either the *PId* or one of these markers. We also calculate the percentage of found publications in MAG, OA, CR, and the combination of the three for each candidate.

The results of this procedure are presented in the following section.

## 4. Results

Overall, we compare the coverage of 2,353,872 publications by 58,335 candidates across 190 RFs. In the following sections we refer to Crossref, OpenAIRE and Microsoft Academic Graph datasets as CR, OA and MAG respectively.

*4.1. Overall coverage by dataset*

Figure 1 shows the overall coverage of candidates' publications in each of the three datasets, as well as their combination. Each data point represents the percentage of publications found in the dataset of interest for a single candidate. Percentages are calculated by taking the number of found publications and dividing this number by the total number of publications in the CV that have relevant publishing metadata. In this diagram, we consider all the candidates who took part in the 2016 - 2018 NSQ sessions, regardless of their RF.

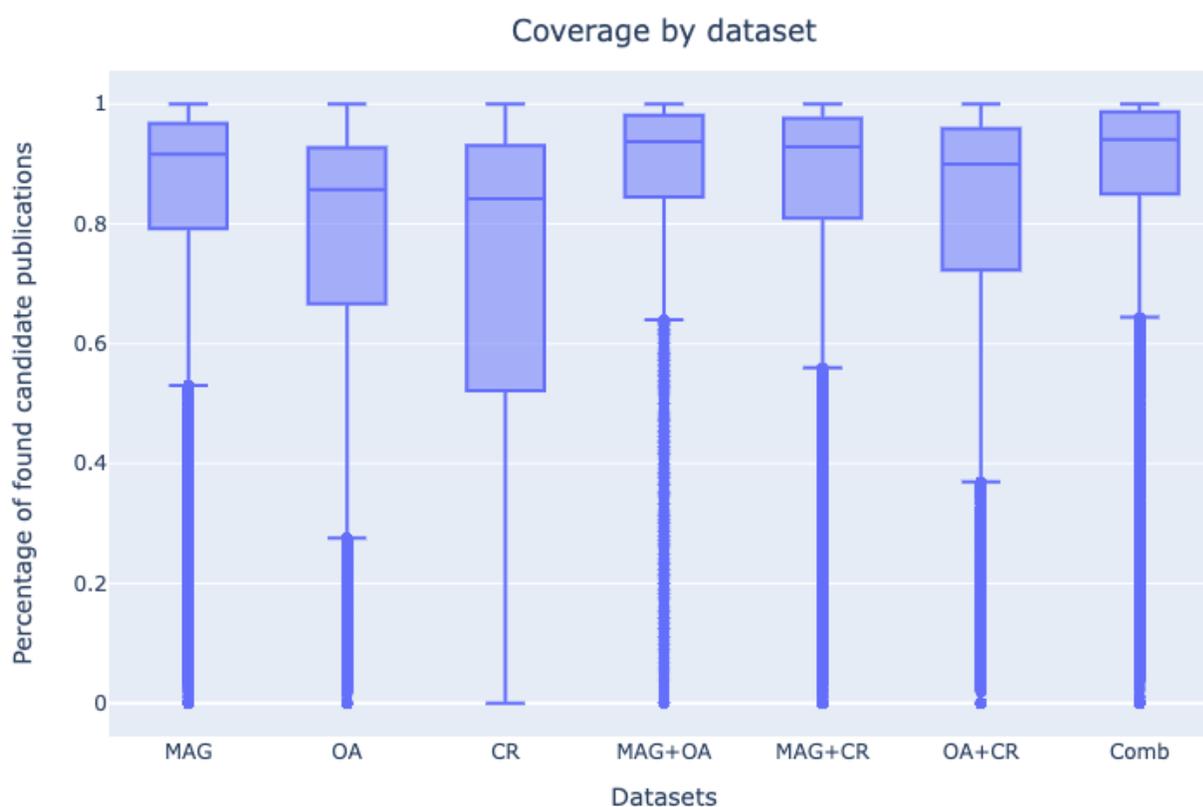

*Figure 1. Overall coverage of candidates' publications in each of the three datasets – Microsoft Academic Graph (MAG), OpenAIRE (OA) and Crossref (CR) – as well as their multiple combinations.*

Overall, all three datasets have very good coverage of candidates' publications. However,

CR's distribution presents a slightly different behaviour: CR's minimum is distinctively lower than that of the other datasets; its first and second quartiles are more spread out; and there are no outliers. This indicates that there are more data points in the lower half of CR's distribution than in the other datasets' distributions. We hypothesize that this phenomenon is due to how metadata is collected to build these datasets and what methods are used for this purpose. MAG is built using web crawling and OA by joining the metadata shared by a network of EU institutions and libraries (including Italian libraries, that collect bibliographic data about all the Italian researchers participating in the NSQ). Whereas, CR is built by its members (i.e. the publishers) and is predominantly DOI-based. Therefore, publications that are not assigned a DOI could be found in MAG and OA, but not in CR.

*4.2. Coverage by citation-based and non-citation-based Disciplines*

Figure 2 displays the coverage of candidates' publications by datasets and field category. As we expected, there is a sharp difference in coverage between candidates who applied to CDs and those who applied to NDs. CDs' distributions are more in line with the overall results shown than NDs' distributions. This phenomenon is caused by the overwhelmingly higher number of candidates in CDs than in NDs – 41,668 applications to CDs and 16,668 applications to NDs – and the disproportionately higher number of publications in CDs than NDs – 1,951,515 publications in CDs, making up for 82% of the overall number of publications, and 402,357 publications in NDs. Indeed, the median number of publications per candidate in CDs is 35, whereas in NDs is 20. As a result, CDs' coverage weighs more heavily in the overall results.

Once again, CR's results are worse than those of MAG and OA, both for CDs and NDs. However, CR's coverage is particularly low in NDs, as its median percentage of found candidate's publication is at 13%.

Furthermore, there is little difference in coverage of CDs between MAG and the combination of the three datasets for CDs, indicating that MAG contains almost all of the open publication data for those disciplines. In addition, there is virtually no difference in coverage of CDs between the combination of MAG and OA and the combination of all three datasets, and between MAG and the combination of MAG and CR. This shows that OA contributes almost all of the additional data not covered in MAG. We find an almost identical pattern in the coverage of NDs. Given this evidence and the demonstrated poor coverage of NDs by CR, we hypothesize that the added data comes from OA.

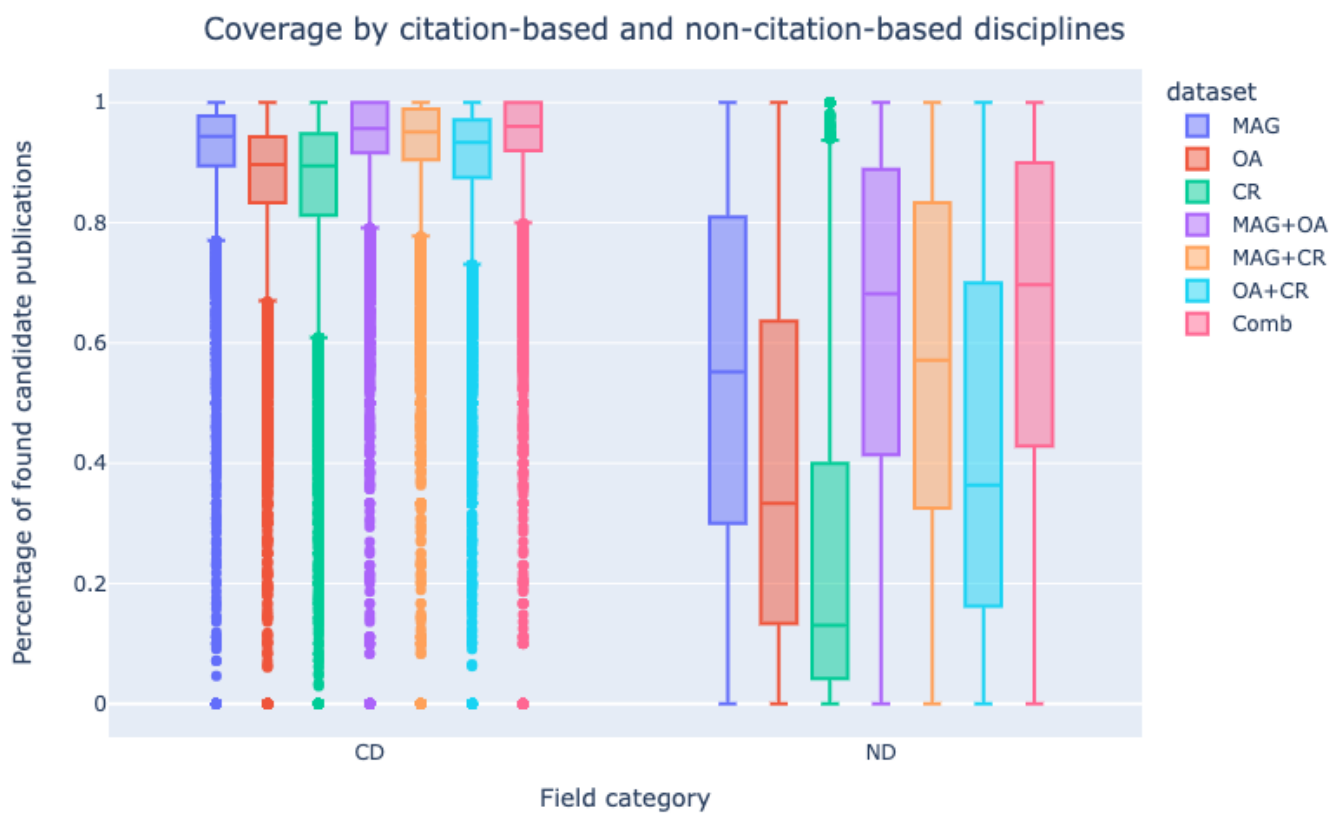

*Figure 2. The coverage of candidates' publications by dataset and field category.*

*4.3. Coverage by datasets and roles*

The diagram in Figure 3 shows the coverage of candidates' publications by dataset and the academic role the candidate applied to (Associate Professor and Full Professor). There is not much difference in coverage between candidates applying for AP and those applying for FP. After all, in the NSQ, candidates are evaluated on the publications published in the last 15 years and in the last 10 years for FP and AP respectively. Therefore, any older publication - that could have affected the results, weighting more for FP who had published more - is not included in the CVs and not considered in this study.

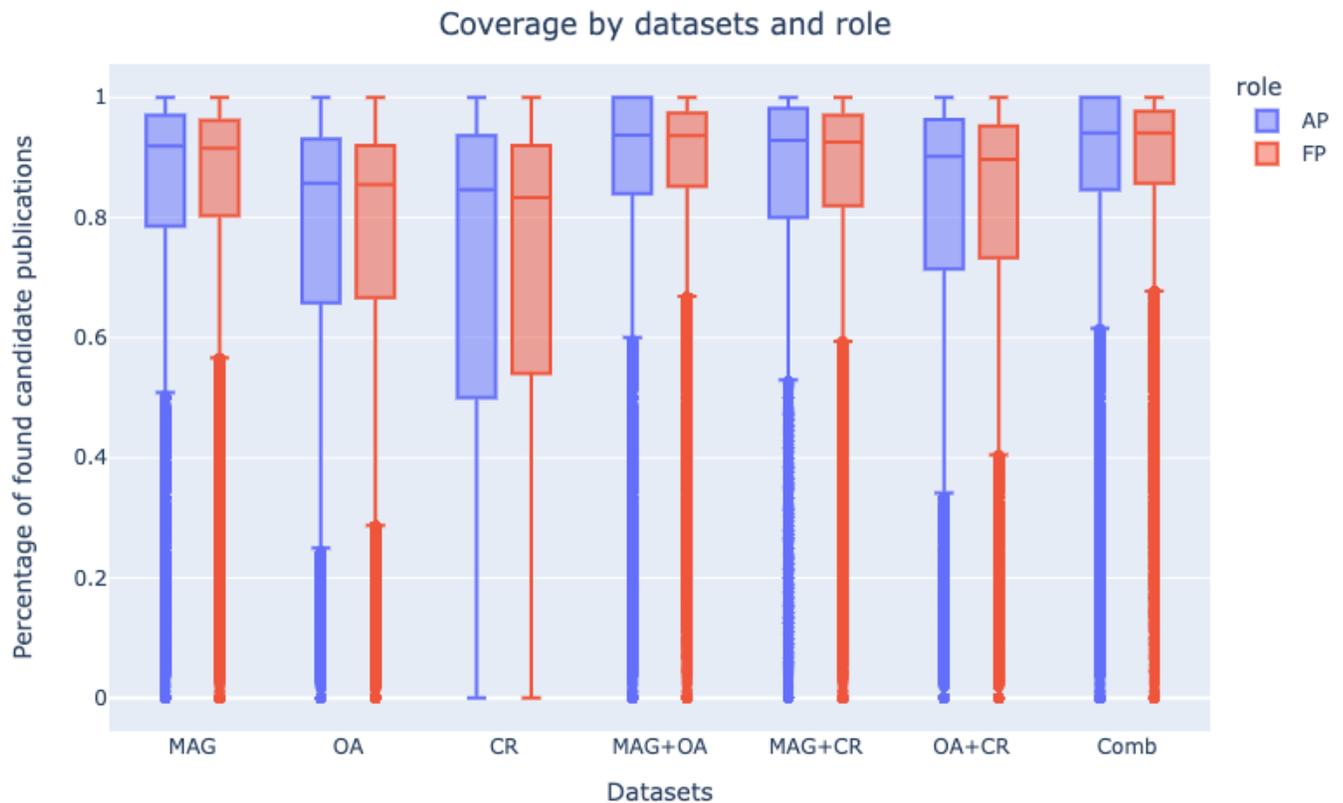

*Figure 3. The coverage of candidates' publications by dataset and academic role the candidate applied to in the NSQ.*

*4.4. Coverage by datasets and scientific areas*

Figure 4 and Figure 5 present the coverage of candidates' publications by dataset and SA (Scientific Area). SAs solely constituted by CDs – 1 through 7 and 9 – all show great coverage results, with tight quartiles and median values that are above 0.85.

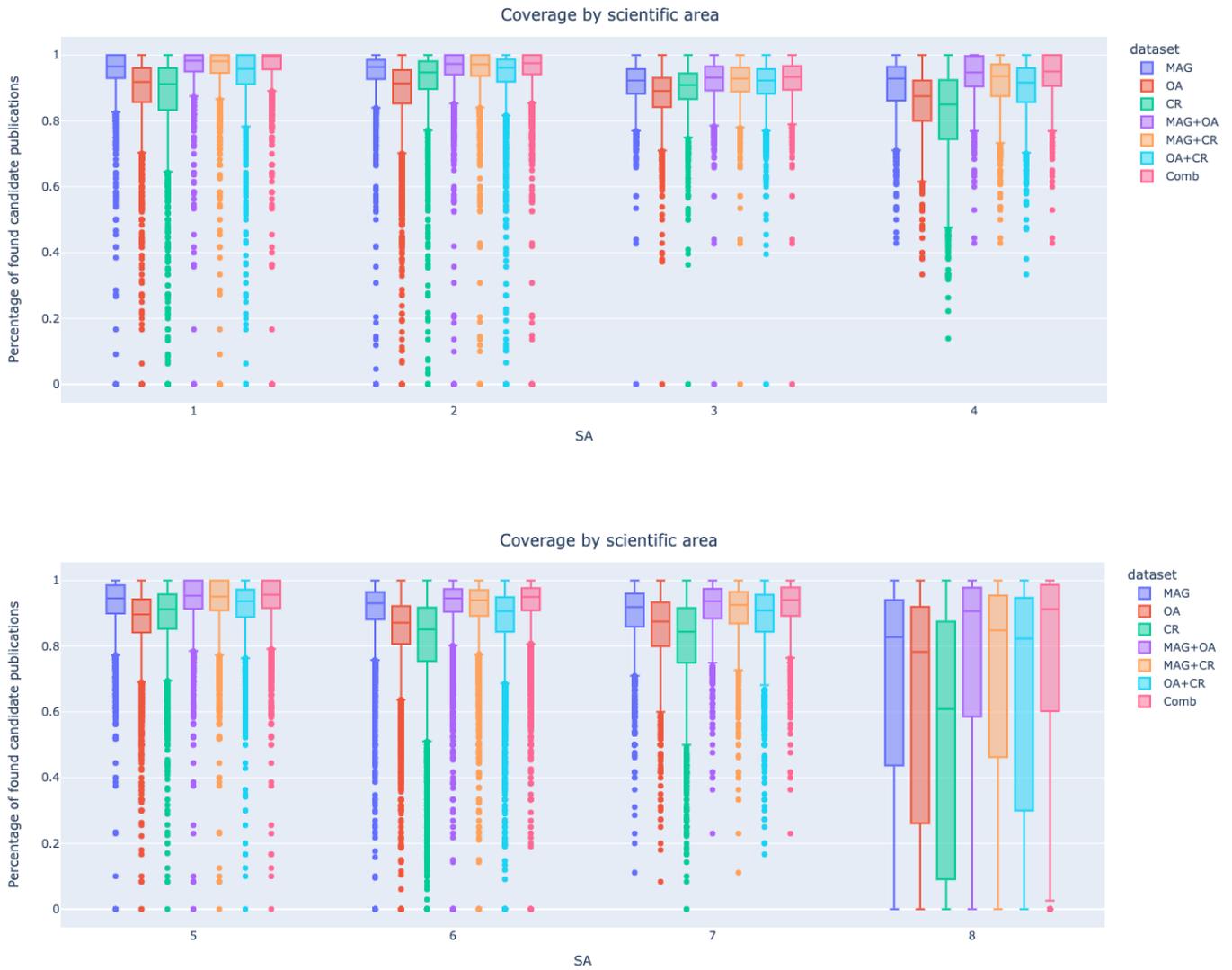

*Figure 4. The coverage of candidates' publications by dataset and SA 1-8.*

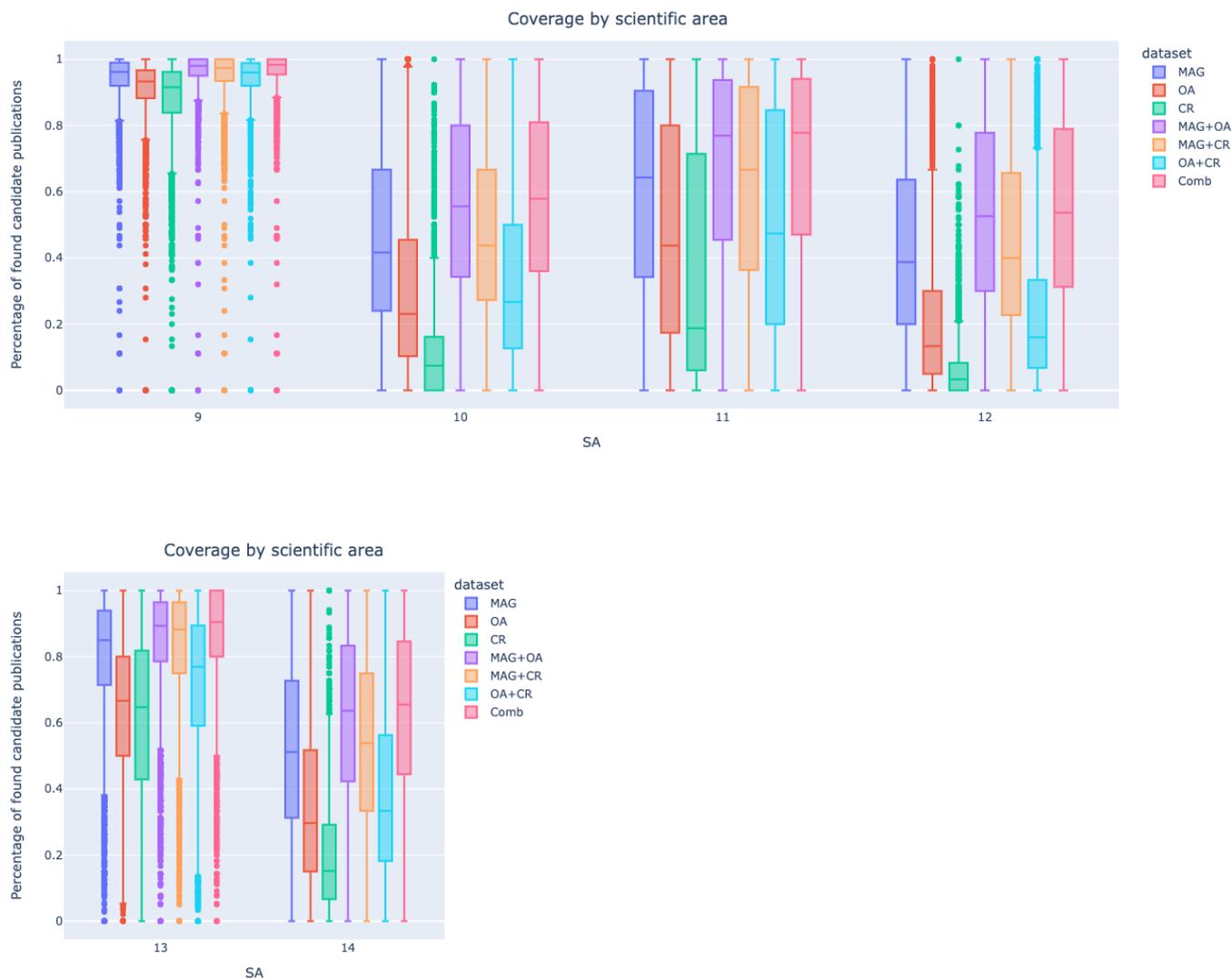

*Figure 5. The coverage of candidates' publications by dataset and SA 9-14.*

SAs solely constituted by NDs – 10 and 12 through 14 – show the worst coverage results, with the exception of SA 13 *Economics and Statistics*. Indeed, 13 presents higher values than 10, 11 and 14, probably caused by its proximity in topic to other SAs only consisting of CDs. However, it also has wider first quartiles than SAs only consisting of CDs, indicating the greater presence of low data points in its distribution.

SAs 8 and 11 are constituted of both CDs and NDs and their distributions are characterized by wide quartiles, indicating the joint presence of high and low data points. It is also worth pointing out CR's poor coverage of 10, 11, 12 and 14, with median values below 20%.

*4.5. Coverage by field in the scientific area with CDs and NDs*

The diagrams in Figure 6 and Figure 7 present coverage of candidates' publications by dataset and RF in mixed SAs (i.e. constituted by CDs and NDs): 8, *Civil Engineering and Architecture*; and 11, *History, Philosophy, Pedagogy and Psychology*. When focusing on the individual RFs inside mixed SAs, the difference in coverage between CDs and NDs clearly emerges. CDs are the RFs with higher values, whereas NDs are the RFs with lower values. 08/D1, *Architectural Design*, presents the worst coverage percentages with median values sharply below 50% for all three datasets, as well as their combination.

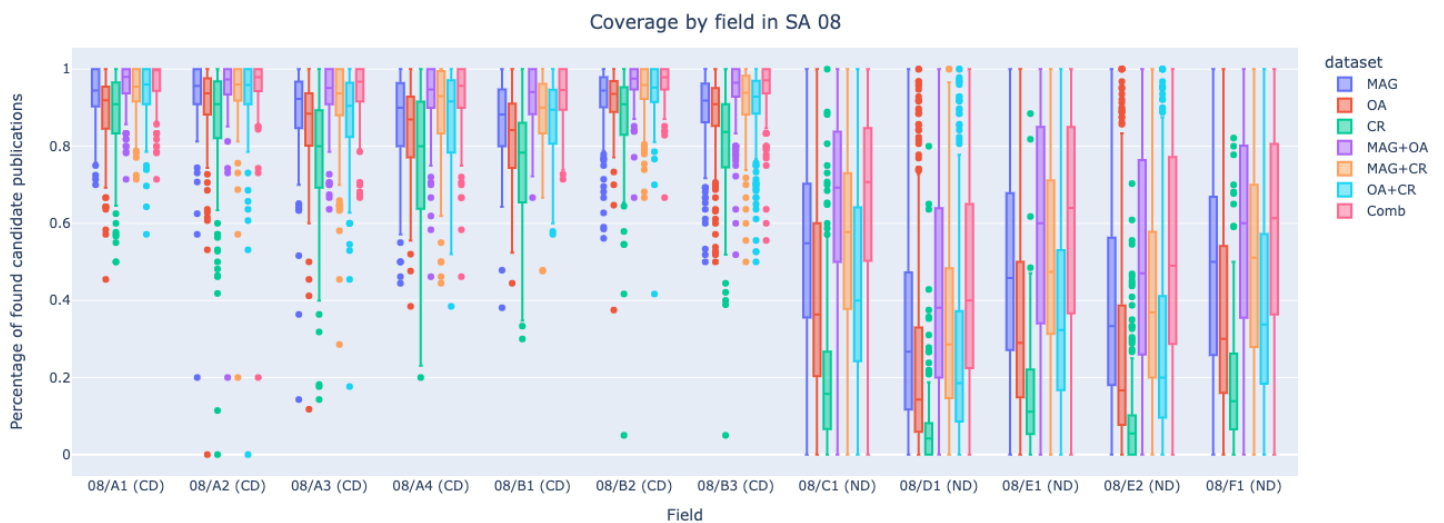

*Figure 6. The coverage of candidates' publications by dataset and RF in SA 8 (Civil Engineering and Architecture).*

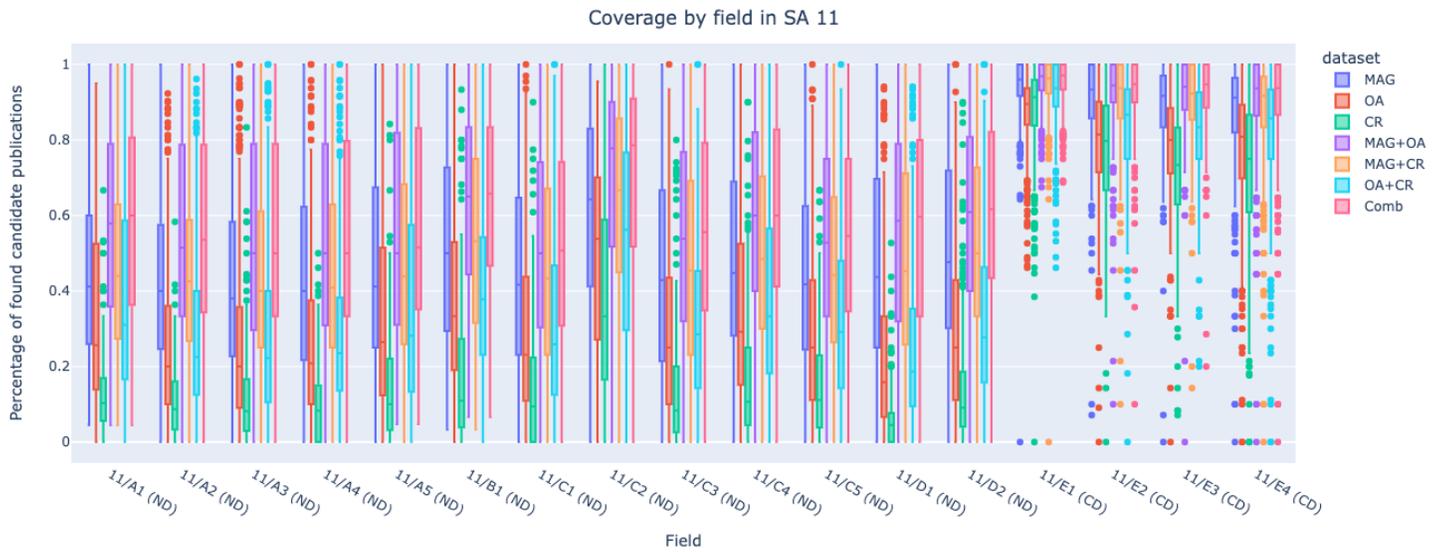

*Figure 7. The coverage of candidates' publications by dataset and RF in SA 11 (History, Philosophy, Pedagogy and Psychology).*

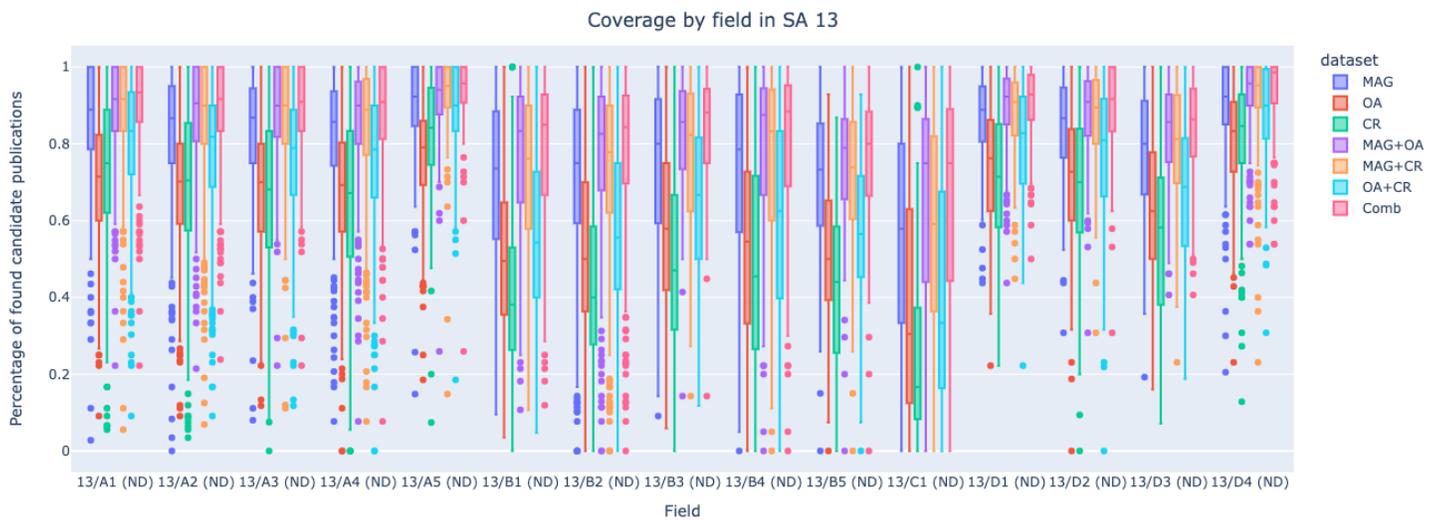

*Figure 8. The coverage of candidates' publications by dataset and RF in SA 13 (Economics and Statistics).*

SA 13, *Economics and Statistics*, is not officially considered a mixed SA by ANVUR –

Open bibliographic data and the Italian NSQ                                                                22indeed it is entirely composed by NDs. However, as shown in Figure 8, it presents similar behaviors and characteristics to mixed SAs: some RFs inside SA 13 have high coverage percentages – 13/A, *Economics*, and 13/D, *Statistics and Mathematical Methods for Decisions* – while others have low coverage percentages – 13/B, *Business Administration and Management*, and 13C, *Economic History*. This diagram casts a light on how publishing and editorial traditions connected to specific topics and disciplines greatly influence the coverage of publications in such fields by open datasets.

## 5. Discussion and conclusions

The main aim of our work was to check what is the coverage, in terms of bibliographic entities, of three of the most prominent and discipline-agnostic open datasets considering a collection of publications mediated by a particular use-case scenario, i.e. the Italian National Scientific Qualification (NSQ). We have shown that the coverage is pretty good in particular for those RFs for which citation data are considered in the evaluation of the NSQ, namely those belonging to SAs from 1 to 7, part of SA 8, SA 9, and part of SA 11. This is reasonable, considering how the NSQ works. Indeed, a candidate applying for a CD recruitment field can present his/her publications if and only if those are listed in Scopus or Web of Science. The list of publications we used matches the data available in the proprietary services. Thus, considering that the coverage for CD as gathered from open datasets is very high (~95%, as shown in Figure 2) we conclude that, in principle, these datasets are ready to be used as sources for evaluation processes in CD-based recruitment fields as an alternative for proprietary bibliographic databases. However, it is worth mentioning that, according to our analysis, they can be used only for identifying the relevant articles a candidate can propose in the application, but we do not have any evidence about their potential use as an alternative for computing the citation-based metrics used in the NSQ, namely the h-index and the citation count.

It is also worth mentioning that, in this work, we have addressed only one step of the assessment process, i.e., the availability of (some) data to perform quantitative measurements. In the context of research assessment exercises, in fact, replacing closed data sources with open ones is an important step to address but it is not substantial for the scholarly community. Indeed, the community urges design research evaluation processes which consider multiple factors – like sources' reputation and strength of the indicators, just to cite a few. These and other aspects might need to be reshaped in such a new context and are under discussion by the community and institutions – e.g., see (Directorate-General for Research and Innovation, European Commission, 2021).

While we claim that, in principle, open datasets can be used in the NSQ when assessing CD-based recruitment fields, the same conclusion does not apply to ND disciplines, where the coverage was lower. Note that, in this case, the candidates certify themselves that the metadata of the publications they present in the NSQ are correct without mandatorily verifying them using external services. Thus, a comparison with these services is much more difficult to perform. Furthermore, we cannot compare the coverage of NDs directly against proprietary bibliographic databases (i.e., Scopus and Web of Science) since we did not have access to their full datasets, even if it could be a good input for a future study. However, by analyzing the data summarized in Figures 4-8, we conclude that there is a clear distinction between the RFs that relate with CDs and those that relate with NDs. Indeed, all the RFs of the first kind (all of that included in SAs 1-7 and 9, plus RFs 08/A1-4, 08/B1-B3, 11/E1-4) are characterized by having very high coverage in the combined dataset (i.e., more than 95%) and at least two out of three datasets with coverage of more than 90%. All the other RFs, which do not comply with these rules, are indeed related to NDs. SA 13 (Figure 8) seems to be a slight exception to this rule: while being labelled by NDs only, it is more blurred due to its intrinsic nature, and it is very close to satisfying the rules mentioned above for some of its RFs.

We can also compare the results of our coverage of CDs against that obtained in other studies on MAG, Scopus and Web of Science. For instance, Visser et al. (2021) show that MAG contains 81% of the publications that are listed in Scopus, computed considering the whole datasets in consideration. This value aligns well with our results, which are even better in favor of MAG, which contains more than 90% of the articles of CDs that are also included in Scopus union Web of Science (as reported in Figure 3). This higher coverage compared with that of Visser et al. can result from:

a) the additional matching of the articles in our collection with Web of Science, that may have brought to a better coverage; and

b) the fact that Italian authors applying to the NSQ for some CDs, being aware that only publications listed in Scopus and Web of Science will be considered, tend to publish more in venues that are indexed in such proprietary bibliographic databases.

Another study by Huang et al. (2020) compared MAG against Scopus and Web of Science using a subset of articles published by authors working in fifteen distinct institutions. Their study shows that MAG covers around 70% and 69% of Scopus and Web of Science publications, respectively. Instead, considering Scopus and Web of Science as a unique dataset with disambiguated publications, MAG covers 67% of the publications in such proprietary services, which is much lower of the coverage (shown in Figure 2) we obtained for CDs and very close to that we have for NDs. In a future study, it could be worth investigating whether the disciplines to which the articles used in the study by Huang et al. (2020) are either CDs or NDs, to check if that low-coverage behavior could be derived from this aspect or is due to other factors.

The ways for correcting data in open and commercial datasets is also worth discussing. Currently, Crossref and MAG do not enable authors to correct wrong metadata of their own publications, while OpenAIRE allows University libraries for validating the metadata they provide to OpenAIRE and for enriching the metadata records with missing or extra information (Manghi et al., 2014). Instead, commercial services usually have curatorial units that react to authors' feedback when provided. This feature is perceived as an added value that can enable authors to have clean data before they are gathered for the evaluation in the NSQ. In principle, this is possible for open datasets as well: the authors could provide similar corrections to the public, thus enabling the teams managing the open datasets their reuse and ingestion and allowing the NSQ commissions to correct possible mistakes in the original data before starting the evaluation procedure.

By analyzing the metadata gathered from the three open sources, we observe that the union of the data in MAG and OA approached that of all the datasets combined, as shown in Figures 1-8. This can mean that the support of Crossref data in our analysis did not add a lot to the other two datasets. However, probably, MAG and OA ingested Crossref data in advance, which can explain why Crossref did not show a big contribution to the overall coverage. In addition, the use of MAG could be perceived as a limit for the replicability of our study using updated data since MAG's discontinuation at the end of 2021. Indeed, it has been recently replaced by OpenAlex (https://openalex.org/) – its first release has been entirely based on the last available snapshot of MAG. However, it is still not clear if OpenAlex future releases will show the same data coverage that Microsoft guaranteed with MAG.

There is an orthogonal aspect of the Italian NSQ that is not addressed in our study, and that will be explored in future works: citation coverage for CDs. Indeed, one of the parameters that are checked by the NSQ is the number of citations all the candidates' publications received in the past. While we claim that the coverage of the open datasets used is good compared with that of Scopus and Web of Science, we cannot affirm the same for citation counts, at least in the context of the NSQ. Indeed, another study we have recently performed (Bologna et al., 2021a) shows that open citation data available in the December 2020 release (OpenCitations, 2020) of OpenCitations' COCI (Heibi et al., 2019) are not yet complete to substitute the data used in the NSQ made available by proprietary services. However, the combined use of several open citation sources – such as the new release of COCI (OpenCitations, 2022), which includes more than 1.29 billion citations, and the additional citation data from the open datasets used in this work and others, such as DataCite (Brase, 2009), and the proved fact that open citations are drastically increasing in these years (Hutchins, 2021) – are encouraging. In addition, a recent update (Martín-Martín, 2021) of a study by Martín-Martín et al. (2021) showed that the coverage of open citation data is approaching parity with those of Web of Science and Scopus. In a future study, we plan to analyze also the coverage of citations in the context of the NSQ, and to compare the results with those available in past studies such as (Visser et al., 2021) and (Martín-Martín et al., 2021).

Finally, it is important to stress one last point. As mentioned above, the NSQ evaluation only takes into account the publications indexed in Web of Science or Scopus for CDs. As shown in the outcomes of our study, there is a key distinction between the coverage of CDs and NDs in the open datasets: they are authoritative for CDs (at least compared with Web of Science and Scopus) and there are a few cases of publications that are relevant for the community but not listed there. The scenario is totally different for NDs disciplines, where a lot of relevant works (e.g. books discussing Humanities and Social Sciences research) are often omitted in commercial repositories. These publications might be instead available in open datasets. Then, it would be interesting to also investigate how much information is missing in Scopus and Web of Science but available in open datasets. This aspect could not be measured so far – since we start from the list of CDs publications selected by the candidates from the commercial datasets only and we could not measure the NDs publications in either Web of Science or Scopus since we did not have access to these commercial indexes – but we plan to explore it with a specific study in the future.

Open bibliographic data and the Italian NSQ                                                                27

Hug, S. E., & Brändle, M. P. (2017). The coverage of Microsoft Academic: Analyzing the publication output of a university. Scientometrics, 113(3), 1551–1571. https://doi.org/10.1007/s11192-017-2535-3

Hutchins, B. I. (2021). A tipping point for open citation data. Quantitative Science Studies, 2(2), 433–437. https://doi.org/10.1162/qss_c_00138

L. 240/2010, Rules concerning the organization of the universities, academic employees and recruitment procedures, empowering the government to foster the quality and efficiency of the university system (Norme in materia di organizzazione delle università, di personale accademico e reclutamento, nonché delega al Governo per incentivare la qualità e l'efficienza del sistema universitario), Gazzetta Ufficiale Serie Generale n.10 del 14/01/2011—Suppl. Ordinario n.11 (2011). https://www.gazzettaufficiale.it/eli/id/2011/01/14/011G0009/sg

Manghi, P., Bolikowski, L., Manold, N., Schirrwagen, J., & Smith, T. (2012). OpenAIREplus: The European Scholarly Communication Data Infrastructure. D-Lib Magazine, 18(9/10). https://doi.org/10.1045/september2012-manghi

Manghi, P., Atzori, C., Bardi, A., Baglioni, M., Schirrwagen, J., Dimitropoulos, H., La Bruzzo, S., Foufoulas, I., Löhden, A., Bäcker, A., Mannocci, A., Horst, M., Jacewicz, P., Czerniak, A., Kiatropoulou, K., Kokogiannaki, A., De Bonis, M., Artini, M., Ottonello, E., … Principe, P. (2021). OpenAIRE Research Graph Dump (3.0). Zenodo. https://doi.org/10.5281/zenodo.4707307

Manghi, P., Artini, M., Atzori, C., Bardi, A., Mannocci, A., La Bruzzo, S., Candela, L., Castelli, D., & Pagano, P. (2014). The D-NET software toolkit: A framework for the realization,

Open bibliographic data and the Italian NSQ 33